# COUPLED LUTTINGER LIQUIDS

H.J. Schulz

*Laboratoire de Physique des Solides, Université Paris–Sud, 91405 Orsay, France*

Many one–dimensional quantum systems, in particular interacting electron and spin systems, can be described a Luttinger liquids. Here, some basic ideas of this picture of one–dimensional systems are briefly reviewed. I then discuss the effect of interchain coupling for a finite number of parallel chains. In the case of spin chains coupled by exchange interactions, the low–energy properties are radically different according to whether the number of coupled chains is even or odd: even number of chains have a gap in the spin excitations, whereas odd numbers of chains are gapless. The effect of interchain tunneling is analyzed for two and three coupled chains of itinerant fermions: for repulsive interactions, the two–chain system is "universally" found to be a d–wave superconductor, with a gap in the spin excitation spectrum. On the other hand, for three chains the ground state depends both on the boundary conditions in the transverse direction and on the strength of the interactions. Weak repulsive interactions in all cases lead to dominant superconducting pairing of d–type. An example of a three–leg spin ladder with a spin gap is proposed. A general scheme to keep track of fermion anticommutation in the bosonization technique is developed.

## I. INTRODUCTION

Much of our current understanding of the physics of interacting electron systems, in particular in metals, is based on Landau's Fermi liquid theory: [1] the properties of the interacting system (the Fermi *liquid*) closely resemble those of a system of noninteracting fermions (the Fermi *gas*). Interactions only lead to finite renormalizations of physical properties (specific heat, spin susceptibility, ...), as well as possibly to symmetry breaking phase transitions. Fundamentally, the properties of a Fermi liquid are all related to the existence of quasiparticle excitations whose lifetime diverges as $1/\omega^2$ when the excitation energy $\omega$ goes to zero. This leads to universal behavior of correlation functions at low energies.

A radically different situation is found in one–dimensional interacting systems: due to the restricted motion along only one direction in space (or equivalently, the not very restrictive conditions on scattering phase space) perturbations easily propagate coherently. This leads in particular to the fact that all the low–energy excitations are collective modes, namely long–wavelength fluctuations of the density (or spin density), and no quasiparticle type elementary excitations exist. Correlation functions then have nonuniversal, interaction–dependent power–law behavior at low energies, and for the case of spin–1/2 fermions the interesting phenomenon of spin–charge separation arises.

In the following section I will review some important results from the theory of spinless interacting fermions in one dimension and their application in particular to spin–chain systems. Both isolated or coupled parallel chains will be considered. Most of these results are well-known, but they will be useful for the understanding of section III, where the case of spin–1/2 fermions will be considered and recent results on the coupled chain problem will be reported. For a general description of the subject, the reader is referred to review articles. [2–4]

## II. THE SPINLESS LUTTINGER LIQUID: SPIN CHAINS

### A. Luttinger model and Luttinger liquid

The basic ingredient of Luttinger liquid theory is the existence of an exactly solvable model of interacting fermions in one dimension, the Luttinger *model*. [5,6] In this model the kinetic energy is essentially that of relativistic one–dimensional fermions, described by the Hamiltonian

$$H_0 = v_F \sum_k \{(k - k_F)a^\dagger_{+,k}a_{+,k} + (-k - k_F)a^\dagger_{-,k}a_{-,k}\} \ , \tag{1}$$

where I use $a_+$–($a_-$–)operators for right–(left–)moving fermions, and $v_F$ is the Fermi velocity. Here the $k$–summation runs from $-\infty$ to $\infty$. In the noninteracting ground state all negative energy states are filled. Note that $k_F$ can be removed by a simple unitary transformation. Interactions between fermions are of the *forward scattering* type, $(k_F; -k_F) \to (k_F; -k_F)$ or $(k_F; k_F) \to (k_F; k_F)$. The Hamiltonian describing these processes takes the form

$$H_{int} = \frac{1}{2L} \sum_q \{2g_2(q)\rho_+(q)\rho_-(-q) + g_4(q)[\rho_+(q)\rho_+(-q) + \rho_-(-q)\rho_-(q)]\} \ . \tag{2}$$

Here, $g_2(q)$ and $g_4(q)$ are the Fourier transforms of a real space interaction potential, $\rho_\pm(q)$ are the Fourier components of the fermion density operator for right–(+) and left–(–)–going fermions, and $L$ is the length of the system.

The model defined by eqs.(1) and (2) can be solved exactly by the bosonization technique. [6] One introduces a scalar boson field $\phi(x)$ and its conjugate momentum density $\Pi(x)$, satisfying canonical boson commutation relations:

$$[\phi(x), \Pi(y)] = i\delta(x - y) \ . \tag{3}$$

$\phi(x)$ is related to the local fermion density via

$$\partial_x \phi(x) = -\pi[\rho(x) - \rho_0] = -\pi[\psi^\dagger_+(x)\psi_+(x) + \psi^\dagger_-(x)\psi_-(x) - \rho_0] \ . \tag{4}$$

Here $\rho_0$ is the average particle density, and the $\psi_\pm$ are standard fermion field operators. In terms of $\phi$ and $\Pi$ the full Hamiltonian becomes

$$H = H_0 + H_{int} = \int dx \left[\frac{\pi u K}{2}\Pi(x)^2 + \frac{u}{2\pi K}(\partial_x \phi)^2\right] \ . \tag{5}$$

This is obviously just the Hamiltonian of an elastic string, with the eigenmodes corresponding to the collective density fluctuations of the fermion liquid. It is important to notice that these collective modes are the only (low–energy) excited states, and that in particular *there are no well–defined single particle excitations*. The parameters in (5) are given by

$$u = [(v_F + g_4/(2\pi))^2 - g_2^2/(2\pi)^2]^{1/2} \ , \ K = \left[\frac{2\pi v_F + g_4 - g_2}{2\pi v_F + g_4 + g_2}\right]^{1/2} \ . \tag{6}$$

The energies of the eigenstates are $\omega(q) = u|q|$, i.e. the parameter $u$ determines the phase velocity of the low–lying excitations.

The final ingredient for the following is the expression for the single–fermion field operators [7,8]

$$\psi_\pm(x) = \frac{1}{\sqrt{L}} \sum_k a_{\pm,k} e^{ikx} = \lim_{\alpha \to 0} \frac{1}{\sqrt{2\pi\alpha}} \eta_\pm \exp\left[\pm ik_F x \mp i\phi(x) + i\theta(x)\right] . \quad (7)$$

Here $\partial_x \theta(x) = \pi \Pi(x)$, $\alpha$ is a short–range cutoff parameter, and the $\eta_\pm$ are Majorana ("real") fermion operators introduced to guarantee proper anticommutation between $\psi_+$ and $\psi_-$. [9] They satisfy the anticommutation relation

$$[\eta_r, \eta_s]_+ = 2\delta_{r,s} , \quad (8)$$

implying in particular $(\eta_r)^2 = 1$. They can be represented by standard (Dirac) fermion operators $c_r$ as $\eta_r = c_r^\dagger + c_r$. Note that there is just one isolated fermionic degree of freedom per branch, and that these degrees of freedom do not appear in the bosonized Hamiltonian.

Using eq.(7) correlation functions can now be easily calculated. One has the general result for ground state expectation values of the Hamiltonian (5)

$$\left\langle e^{i(m\phi(x,\tau)+n\theta(x,\tau))} e^{-i(m\phi(0,0)+n\theta(0,0))} \right\rangle \propto \cos[mn\chi(x,\tau)](x^2 + u^2\tau^2)^{-\eta(m,n)/2} ,$$
$$\eta(m,n) = \frac{1}{2}(m^2 K + n^2/K) . \quad (9)$$

Here $\chi(x,\tau)$ is the angle between the vector $(x, u\tau)$ and the $x$–axis, and $\tau$ is the Matsubara imaginary time. Thus, quite generally, correlation function decay asymptotically with *nonuniversal power laws* which exponents depending on the strength of the interaction via the coefficient $K$, eq.(6). After Fourier transformation this of course gives nonuniversal power laws at low energies and long wavelengths. As an example consider the pairing correlation function

$$\langle \psi_+^\dagger(x,\tau)\psi_-^\dagger(x,\tau)\psi_-(0,0)\psi_+(0,0)\rangle \approx \langle e^{2i\theta(x,\tau)} e^{-2i\theta(0,0)}\rangle \propto (x^2 + u^2\tau^2)^{-1/K} . \quad (10)$$

As expected in a one–dimensional model, this correlation function never shows long–range order, however, for attractive interaction (corresponding to $K > 1$) where standard BCS theory would predict long–range order, the decay is slower than in the noninteracting case, indicating enhanced pairing correlations. Similar results can be found for a large variety of other correlation functions.

The above is exact for the Luttinger *model*. On the other hand, one would expect that a model given by eqs.(1) and (2) should also be appropriate for *the low–energy properties* of, for example, models for interacting fermions moving on a lattice or with nonlinear energy dispersion. This is certainly correct in the weak–interaction limit where all the low–energy physics is dominated by processes near the Fermi level and the structure of states far from the Fermi level is largely irrelevant. By continuity as long as no phase transition occurs as a function of increasing interaction strength, eq.(5) then also is the effective low–energy Hamiltonian for stronger interactions. The coefficients $u$ and $K$ of course depend in a more or less complicated fashion on the interaction, but the basic fact that all the low–energy physics is determined entirely by $u$ and $K$ remains. In particular, scaling relations between exponents of different correlation functions are always the same. This reasoning can be made quantitative using renormalization group arguments to show that (almost) all the effects neglected (band curvature, finite cutoffs, momentum dependence of the interaction,...) are irrelevant and therefore only can lead to quantitative renormalizations of $u$ and $K$. [10] Physical systems to which this kind of description applies include, apart from interacting electrons, also one–dimensional spin and interacting boson systems, as well a the edge states of quantum Hall devices. These system, where the Hamiltonian is correct for low–energy properties but in general not for high–energies

(of order of the bandwidth or higher), are now referred to as *Luttinger liquids*. [10] In all cases, the coefficients $u$ and $K$, sometimes appropriately generalized to include the effect of electron spin, determine low–energy correlation functions, and therefore transport properties, the effect of extra perturbations like random potentials, and low energy spectroscopic properties.

### B. Spin chain problems

A particularly interesting example of the use of the Luttinger liquid concept is provided by the spin–1/2 antiferromagnetic chain, described by the spin Hamiltonian

$$H = \sum_{i=1}^{L}(S_i^x S_{i+1}^x + S_i^y S_{i+1}^y + \Delta S_i^z S_{i+1}^z) \; . \tag{11}$$

Here $\boldsymbol{S}_i = (S_i^x, S_i^y, S_i^z)$ is a spin–1/2 operator acting on site $i$, $\Delta$ is an anisotropy parameter that allows one to treat the antiferromagnetic ($\Delta = 1$), the ferromagnetic ($\Delta = -1$), and general anisotropic cases. The model has been solved exactly by Bethe's celebrated "ansatz", [11,12] but many of its properties have remained difficult to describe until much later.

The Luttinger liquid concept in fact is very useful for describing asymptotics of correlation functions for this model. One starts by the Jordan–Wigner transformation [13] from spin to fermion operators $(a_i, a_i^\dagger)$:

$$S_i^+ = S_i^x + i S_i^y = a_i^\dagger \exp\left[i\pi \sum_{j=1}^{i-1} a_j^\dagger a_j\right] \;,\; S_i^z = a_i^\dagger a_i - \frac{1}{2} \; . \tag{12}$$

In the fermionic language, the $S_i^+ S_{i+1}^-$ interaction leads to a nearest neighbor hopping term, and $S_i^z S_{i+1}^z$ leads to a nearest neighbor interaction. Following the logic outlined above, one then finds eq.(5) as the effective low–energy Hamiltonian, with in particular $K = 1$ for $\Delta = 0$. The dominant (i.e. most slowly decaying) contributions to the spin operators are

$$S_i^+ \approx (-1)^i e^{-i\theta(ia)} \quad,\quad S_i^z \approx (-1)^i \cos(2\phi(ia)) \;, \tag{13}$$

where $a$ is the lattice constant. From eq.(9) the longitudinal and transverse spin–spin correlation functions then decay as $x^{-2K}$ and $x^{-1/(2K)}$, respectively. This immediately implies that for the isotropic antiferromagnet ($\Delta = 1$), where spin rotation invariance imposes equality of longitudinal and transverse correlations, one has

$$\langle \boldsymbol{S}_j(\tau) \cdot \boldsymbol{S}_0 \rangle \approx (-1)^j \frac{\ln^{1/2}(x^2 + (u\tau)^2)}{\sqrt{x^2 + (u\tau)^2}} \quad,\quad x = ja \; . \tag{14}$$

Here the logarithmic factor [14–16] comes from a marginally irrelevant umklapp operator [17] ($\propto \cos(4\phi)$ in bosonic language) which has been neglected up to here. We thus have $K = 1/2$ for the isotropic antiferromagnet. Eq.(14) and its generalization to finite temperature can be used to obtain an analytic expression for the magnetic neutron scattering cross section [18] which is found to be in excellent agreement with experiment. [19,20]

For general values of $\Delta$ (obeying $|\Delta| \leq 1$) $K$ can be determined noting that in eq.(5) the coefficient $u/K$ determines the inverse susceptibility, which can be obtained from the exact Bethe Ansatz solution of the spin model. One then obtains [21,17]

$$K = \frac{\pi}{2\arccos(-\Delta)} \quad,\quad u = \frac{\pi\sqrt{1-\Delta^2}}{2\arccos\Delta} \; . \tag{15}$$

## C. Coupled spin chains

Investigating models of coupled parallel chains is of interest for a number of reasons: (i) *quasi*–one–dimensional antiferromagnets always have some form of interchain coupling, usually leading to three–dimensional ordering at sufficiently low temperatures; [22–24] (ii) there is a number of "spin–ladder" compounds containing a small number of coupled chains; [25,26] (iii) coupled spin–1/2 chain models can be used to describe higher spin quantum numbers. [27]

Consider $N$ coupled spin–1/2 chains with spin degrees of freedom $\boldsymbol{S}_j$, $j = 1..N$, described by the Hamiltonian

$$H = \sum_{j=1}^N H(\boldsymbol{S}_j) + \sum_{j<k} \lambda_{jk} H_c(\boldsymbol{S}_j, \boldsymbol{S}_k) \; , \quad H_c(\boldsymbol{S}_j, \boldsymbol{S}_k) = \sum_i \boldsymbol{S}_{j,i} \cdot \boldsymbol{S}_{k,i} \; , \qquad (16)$$

where $i$ labels sites along the chains, $j,k$ label the chains, and $H(\boldsymbol{S}_j)$ is of the form (11). For $\lambda_{jk} \equiv -1$ the ground state of this model is exactly that of the spin–$N/2$ chain, each site being in a state of total spin $N/2$, [27] but the model can be considered for general $\lambda$, both ferromagnetic ($\lambda < 0$) and antiferromagnetic ($\lambda > 0$). Performing now the Jordan–Wigner transformation for the $\boldsymbol{S}_j$ separately and going to the boson representation the Hamiltonian becomes [18]

$$\begin{aligned} H &= \int dx \left[ \frac{\pi u K}{2} \boldsymbol{\Pi}(x)^2 + \frac{u}{2\pi K} (\partial_x \boldsymbol{\phi})^2 + \frac{u}{2\pi \bar{K}} (\partial_x \bar{\phi})^2 \right] \\ &+ \frac{1}{(\pi \alpha)^2} \sum_{j<k} \int dx \{ \lambda_{1,jk} \cos(2(\phi_j + \phi_k)) + \lambda_{2,jk} \cos(2(\phi_j - \phi_k)) + \lambda_{3,jk} \cos(\theta_j - \theta_k) \} \end{aligned} \qquad (17)$$

Here $\boldsymbol{\phi} = (\phi_1, \phi_2, .., \phi_N)$, $\bar{\phi} = \sum_j \phi_j / \sqrt{N}$, the constants $u, K, \bar{K}$ all depend on the different constants in the original Hamiltonian, and the $\lambda_{1,2,3;jk}$ are all proportional to $\lambda_{jk}$.

Elementary power counting, using the result eq.(14) for the spin correlations, shows that the coupling term $H_c$ always is a strongly relevant perturbation. Consequently, an explicit renormalization group calculation [18] shows that either $\lambda_2$ or $\lambda_3$ always scale to strong coupling, i.e. the "relative" degrees of freedom $\phi_j - \phi_k$ all acquire a gap. In particular, for not too strong anisotropy, the $\lambda_3$ operator dominates, giving rise to long–range order in the $\theta_j - \theta_k$ and correspondingly exponential decay of the $\phi_j - \phi_k$ correlations. Integrating out these massive degrees of freedom an effective Hamiltonian for the "global" $\bar{\phi}$ mode is found:

$$H = \frac{u}{2} \int \left\{ \pi K \bar{\Pi}(x)^2 + \frac{1}{\pi K} (\partial_x \bar{\phi})^2 + g \cos(\mu \sqrt{N} \bar{\phi}) - \frac{\sqrt{N}}{\pi} h \partial_x \bar{\phi} \right\} \; , \qquad (18)$$

where $\mu = 2$ for even $N$ and $\mu = 4$ for odd $N$, the coefficients $u, K, g$ are renormalized, and $h$ is an external magnetic field applied along the $z$ direction. Similarly, the leading contribution to spin correlations comes from the operators

$$S^+(x) \propto e^{i\pi x} e^{-i\bar{\theta}/\sqrt{N}} \; , \quad S^z(x) \propto e^{i\pi x} \cos(2\sqrt{N} \bar{\phi}) \; , \qquad (19)$$

where the second equation applies to odd $N$ only.

From this a number of important conclusions can be drawn. [18] We first notice that massless excitations and the corresponding slow algebraic decay of correlation functions are only possible if the cos term in eq.(18) is irrelevant. Moreover from eqs.(19) and (9) it follows that spin correlation functions are isotropic only if $K = 1/(2N)$, implying a decay as $1/x$, as in the $S = 1/2$ case. For the case of odd $N$ (equivalently, for ferromagnetic $\lambda$, for half–odd–integer S) this is indeed the correct behavior: for $\mu = 4$ the cos term is marginally irrelevant. Thus both antiferromagnetic spin chains with half–odd–integer S and for odd numbers of coupled $S = 1/2$ chains massless behavior is predicted, with correlations asymptotically decaying like those of a spin–1/2 chain. [18] There is both numerical [28–31] and experimental [25] evidence that this is correct.

On the other hand, for even $N$ (equivalently, integer $S$) the cos term in eq.(18) is strongly relevant and therefore generates a gap $\Delta_s$ in the spin excitations. For the integer–$S$ spin chains this is the well–known and verified Haldane prediction [32], but there also is a gap for any even number of coupled chains. [18,33] The gap implies exponential decay of spin correlations. Numerical [30,31] and, at least for $N = 2$, experimental work [25] confirms this picture. Another prediction, again valid both for integer–$S$ antiferromagnets and even numbers of coupled chains, concerns the effect of an applied magnetic field: [18,34,35] as long as the field is smaller than a critical field $h_c \propto \Delta_s$, the ground state is unchanged and has zero magnetization. However, beyond $h_c$ the magnetization is expected to increase as $M \propto \sqrt{h - h_c}$. Experiments on $S = 1$ antiferromagnetic chains confirm this prediction. [36]

A natural question left open by the above considerations concerns quasi–one–dimensional antiferromagnets like KCuF$_3$ [22,23] or Sr$_2$CuO$_3$, [24,37] which can be considered as the $N \to \infty$ limit of the above model. One clearly expects (and observes) true antiferromagnetic order at sufficiently low temperatures, at first sight in contradiction both with the exponential decay of spin correlation predicted for even $N$ and the universal $1/x$ law for odd $N$. However, one should note that on the one hand the correlation length of the even–$N$ systems is expected to increase quickly with increasing $N$, and that on the other hand the $1/x$ correlation law of the odd–$N$ systems also is expected to be only valid beyond a correlation length $\xi(N)$ which increase with $N$. In the thermodynamic limit $N \to \infty$ this then is perfectly consistent with the existence of long–range order. Theoretical treatments of magnetic order in quasi–one–dimensional antiferromagnets can be based on a mean-field treatment of the interchain interaction [38,39] which gives quantitative predictions for systems like KCuF$_3$ or Sr$_2$CuO$_3$.

### III. LUTTINGER LIQUID WITH SPIN: (QUASI–)ONE–DIMENSIONAL CONDUCTORS

#### A. One chain...

The generalization to fermions with spin of the method of the previous section is relatively straightforward: [2] one introduces one $\phi$–field for each spin orientation, and then uses the charge–($\rho$) and spin–($\sigma$) fields $\phi_{\rho,\sigma} = (\phi_\uparrow \pm \phi_\downarrow)/\sqrt{2}$. The dynamics of theses fields is governed by separate charge– and spin–Hamiltonians, each of the form (5), with constants $u_{\rho,\sigma}, K_{\rho,\sigma}$. In addition, backward scattering leads to the existence of an extra term $g_1 \cos(\sqrt{8}\phi_\sigma)$. For repulsive interactions ($g_1 > 0$) this term renormalizes to zero, and I will concentrate on this case in the following. By arguments similar to those used for the spin chain problem, spin rotation invariance requires $K_\sigma = 1$, and then there are three interaction dependent constants $u_\rho, u_\sigma, K_\rho$ left in the effective low–energy Hamiltonian. These can be determined either perturbatively, giving rise to expressions like eq.(6), or can be determined from exact solutions and numerical finite–size diagonalization. [40,41] For a noninteracting system $u_\rho = u_\sigma = v_F$, $K_\rho = 1$.

A remarkable feature is the separation of spin– and charge degrees of freedom: [2] the Hamiltonian is a sum of spin– and charge part, and, from eq.(7) operators and therefore correlation functions become products of a spin– and a charge part. As spin and charge excitations propagate with different velocities, this gives rise to the interesting phenomenon of *spin–charge separation*, leading for example to a complicated double–peak structure in the single–particle spectral density [42,43]instead of the single quasiparticle peak of a Fermi liquid. Charge and

spin–modes propagating at different velocities have indeed been observed in experiments on quantum wires. [44,45] Of course, as in the spinless case, no broken–symmetry ground states occur, but fluctuations of various types of order are enhanced. In particular, for purely repulsive interactions, antiferromagnetic (equivalently, spin density wave) fluctuations dominate.

### B. ...two chains...

It is clearly of interest to see what happens to the peculiar one–dimensional behavior when one puts chains in parallel. This question is of relevance for the understanding of quasi–one–dimensional conductors, doped spin ladders, few–channel quantum wires, and generally for the understanding of possible non–Fermi–liquid behavior and correlation–induced super-conductivity in higher–dimensional solids. Of particular interest is the effect of an interchain single–particle tunneling term of the form

$$H_{ij} = -t_\perp \int dx (\psi^\dagger_{rsi} \psi_{rsj} + h.c.) \ ,  \qquad (20)$$

where $\psi_{rsj}$ is the fermion field operator for right ($r = +$) or left ($r = -$) going particles of spin $s$ on chain $i$. Simple scaling arguments [46] lead to the "phase diagram" shown in fig.1. The dashed line represents the crossover below which single–particle tunneling becomes strongly relevant and below which one thus expects Fermi liquid like behavior (an alternative interpretation is due to Anderson [47,48]). The full lines indicate where two–particle or particle–hole hopping becomes relevant. The most plausible interpretation is that this is the temperature where three–dimensional long–range order of some type sets in. A more detailed discussion of this in quasi–one–dimensional systems (a thermodynamically large number of chains) has been given in this conference. [49,50]

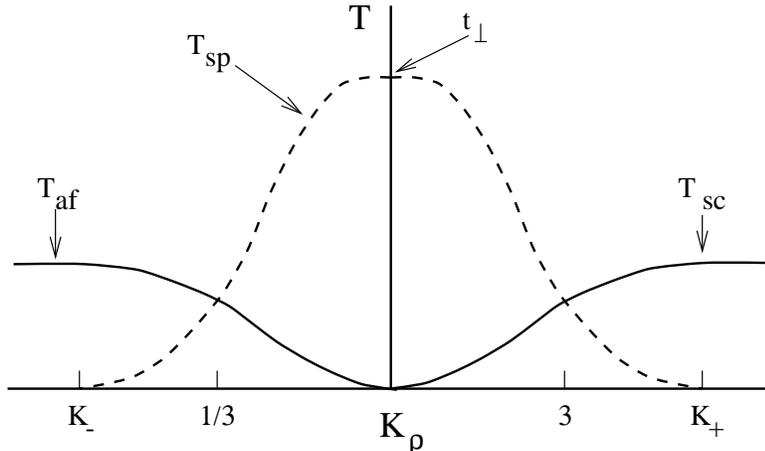

FIG. 1. Qualitative phase diagram in the temperature–$K_\rho$ plane for Luttinger liquids coupled by interchain hopping. $K_\pm = 3 \pm \sqrt{8}$.

As a model for potential doped spin–ladder systems, as well as a first step towards a many–chain system, one can study the two–chain case. [51–53] The tunneling term, eq.(20), then leads to a splitting of the single–particle bands into symmetric and antisymmetric combinations which I label by transverse wavenumbers $k_\perp = 0, \pi$. Now each $k_\perp$ mode can be bosonized separately. Introducing the linear combinations $\phi_{\nu\pm} = (\phi_{\nu 0} \pm \phi_{\nu\pi})/\sqrt{2}$ ($\nu = \rho, \sigma$) the Hamiltonian (including $t_\perp$) then takes the form

$$H = H_0 + H_{int,1} + H_{int,2} \ , \quad H_0 = \frac{\pi v_F}{2} \sum_{\substack{\nu=\rho,\sigma \\ \alpha=\pm}} \int dx \left[ \Pi^2_{\nu\alpha} + \frac{1}{\pi^2}(\partial_x \phi_{\nu\alpha})^2 \right]$$

$$H_{int,1} = -\frac{g_1}{4}\int dx \left[\frac{1}{\pi^2}\{(\partial_x\phi_{\rho+})^2 + (\partial_x\phi_{\sigma+})^2\} - \Pi^2_{\rho+} - \Pi^2_{\sigma+}\right]$$

$$+ \frac{g_1}{2(\pi\alpha)^2}\int dx\{\cos 2\phi_{\sigma+}(\cos 2\theta_{\rho-} + \cos 2\phi_{\sigma-} - \cos 2\theta_{\sigma-}) - \cos 2\theta_{\rho-}\cos 2\theta_{\sigma-}\}$$

$$H_{int,2} = \frac{1}{4}\int dx \sum_{\gamma=\pm} g^{(2)}_\gamma \left[\frac{1}{\pi^2}(\partial_x\phi_{\rho\gamma})^2 - \Pi^2_{\rho\gamma}\right]$$

$$+ \frac{g^{(2)}_{00\pi\pi}}{2(\pi\alpha)^2}\int dx \cos 2\theta_{\rho-}(\cos 2\phi_{\sigma-} + \cos 2\theta_{\sigma-}) \ . \tag{21}$$

Here $g^{(2)}_\gamma = g^{(2)}_{0000} + \gamma g^{(2)}_{0\pi\pi 0}$, and $g^{(i)}_{abcd}$ is the coupling constant for an interaction scattering two particles from $k_\perp$-states $(a,b)$ into $(d,c)$. The signs of the different interaction terms in eq.(21) have been determined following the reasoning explained in the appendix. Here I consider a case where there is only intrachain interaction, implying that all *bare* coupling constants are independent of $k_\perp$.

For the pure forward scattering model ($g_1 \equiv 0$) the only nonlinear interaction ($g^{(2)}_{00\pi\pi}$) scales to infinity, leading to a gap in the $(\rho-)$ modes and in *half* of the $(\sigma-)$ modes. The remaining $(\sigma-)$ modes are protected by the duality symmetry under $\phi_{\sigma-} \leftrightarrow \theta_{\sigma-}$, the $(\sigma-)$ sector is in fact a critical point of the Ising type for $g_1 \equiv 0$. [53] For repulsive interactions here the dominant fluctuations are of CDW type, with decay proportional to $r^{-(3+2K_\rho)/4}$, and $K^2_\rho = (\pi v_F - g_2 + g_1/2)/(\pi v_F + g_2 - g_1/2)$. This state can be label by he number of massless modes as $C1S1\frac{1}{2}$, where quite generally $CnSm$ denotes a state with $n$ massless charge and $m$ massless spin modes. [54]

For nonzero $g_1$ all interactions scale to strong coupling, only the total charge mode remains massless ($C1S0$), reflecting the translational invariance of the system, and all spin excitations have a gap. The physics in this regime can be determined looking for the semiclassical minima of the different cos terms in eq.(21). One then finds that the CDW correlations now decay exponentially, and for the interesting case $g_1 > 0, g_2 > g_1/2$, corresponding to purely repulsive interaction, the strongest fluctuations are now of "d–type" superconducting pairing, [51,53] with decay as $r^{-1/(2K_\rho)}$. Labeling this state as "d–type" seems appropriate because the pairing amplitudes at $k_\perp = 0$ and $\pi$ intervene with opposite sign. In real space, this corresponds to pairs formed by two fermions on the two different chains. Note that even for weak interactions where $K_\rho \to 1$ this decay is very slow. The $4k_F$ component of the density correlations also has a power law decay, however with an exponent $2K_\rho$, much bigger than the SCd exponent. [53,54] The full phase diagram in the $g_1$–$g_2$ plane is shown in fig.2. For $g_1 < 0$ the diagram is identical to the single–chain case, however, for $g_1 > 0$ the behavior is changed dramatically, and in particular superconductivity is predicted for repulsive interactions, for example for the Hubbard model which would be represented as $g_1 = g_2$ in the present language.

Remarkably, results basically identical to this weak coupling analysis can be obtained assuming strong repulsive interactions in the individual chains, so that one is for example in the regime where interchain electron–hole pair hopping is more relevant than single particle tunneling, $K_\rho < 1/3$ in fig.1). It is then more appropriate to bosonize the degrees of freedom of individual chains, rather than working in $k_\perp$–space. Under renormalization one then generates an interchain interaction of the form [46,55]

$$H_{jk} = \frac{1}{(2\pi\alpha)^2}\cos(\sqrt{2}(\phi_{\rho j} - \phi_{\rho k}))$$

$$\times \left\{J_\perp[\cos(\sqrt{2}(\theta_{\sigma j} - \theta_{\sigma k})) + \cos(\sqrt{2}\phi_{\sigma j})\cos(\sqrt{2}\phi_{\sigma k})] + V\sin(\sqrt{2}\phi_{\sigma j})\sin(\sqrt{2}\phi_{\sigma k})\right\} \tag{22}$$

where $j$, $k$ now are chain indices. Remarkably, this term leads to properties identical to those found in weak coupling. [53] First, for $g_1 = 0$ one has $V = J_\perp$. Then $H_{jk}$ is invariant under the duality $\phi_{\sigma j} \leftrightarrow \theta_{\sigma j}$, and one has an Ising critical theory. Secondly, for $g_1 > 0$ one finds $V \neq J_\perp$. Then the fields $\phi_{\rho 1} - \phi_{\rho 2}$, $\phi_{\sigma 1} + \phi_{\sigma 2}$, and $\theta_{\sigma 1} - \theta_{\sigma 2}$ become long–range ordered and one is in a $C1S0$ state. Power law correlations again exist for SCd and $4k_F$ density fluctuations, with the same scaling relation between the two exponents as in weak coupling. However, because now $K_\rho < 1/3$ the $4k_F$ density fluctuations actually dominate. The equivalent results in the weak and strong coupling regime very strongly suggest that *the two–chain model is in the same phase for weak and strong repulsion*. This point is further supported by considering the "t–J ladder model" for strong interchain exchange. [56]

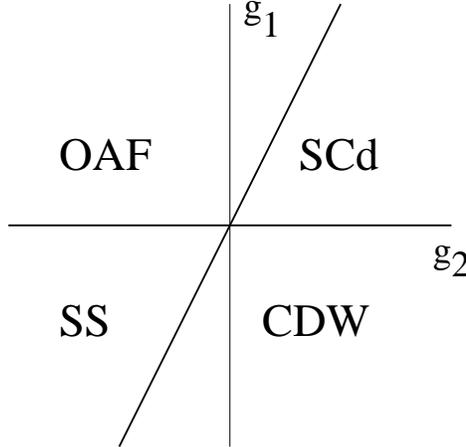

FIG. 2. Phase diagram of the two–chain model. The different dominating fluctuations are: SCd: "d–type" pairing; OAF: orbital antiferromagnetism; SS: "s–type" pairing; CDW: charge density wave. The critical lines $g_1 = 0$ and $g_1 = 2g_2$ are of Ising type.

Numerical work on the two–chain model is in agreement with the existence of d–type pairing, [57–61] the evidence for the $4k_F$ density fluctuations is however inconclusive. [62] Concerning experimental observation, one should notice that the SCd state becomes localized by weak disorder. [63]

### C. ...three chains

Given the differences between even and odd numbers of parallel spin chains discussed in sec. II C, it is clearly of interest to see what happens when one passes from two to three (and eventually more) chains. For the three–chain case it is important to consider boundary conditions in the transverse direction carefully: periodic boundary conditions are frustrating, and I therefore consider both periodic and open boundary conditions. The corresponding geometries and electronic structures near the Fermi energy are shown in fig.3.

In the periodic case, there are 12 different coupling constants implying particles at the Fermi energy. These interactions *do not* satisfy the criteria for bosonization developed in the appendix. However, a one–loop renormalization group calculation shows that certain of these interactions scale to strong coupling, whereas others remain finite. It turns out that the interactions that do scale to strong coupling do also satisfy the criteria of the appendix, i.e. following the same logic as in the two–chain case the strong coupling regime can be understood based on a bosonized picture. Specifically, I then find: (i) if only forward scattering is concerned, one is at an Ising critical point similar to the two–chain case, in a $C1S2\frac{1}{2}$ state. Both CDW and SDW correlations decay exponentially; (ii) upon introducing backward scattering, all the massless spin modes disappear ($C1S0$); (iii) for $g_1 > 0$ and $g_1 < 2g_2$ the dominant fluctuations are of "$d_{xy}$"–type, where by this I mean that the pairing amplitude has opposite signs ant $k_\perp = \pm 2\pi/3$ and vanishes at $k_\perp = 0$; for $g_1 > 0$ and $g_1 > 2g_2$ the pairing amplitude has the same sign at all $k_\perp$. In both cases, the decay exponent is $1/(3K_\rho)$, with $K_\rho$ given by an expression analogous to the two–chain case. We note that at least for two and three chains, the exponent takes the form $1/(NK_\rho)$, with $N$ the number of chains. For arbitrary $N$ this form can be obtained from a microscopically derived Ginzburg–Landau description.

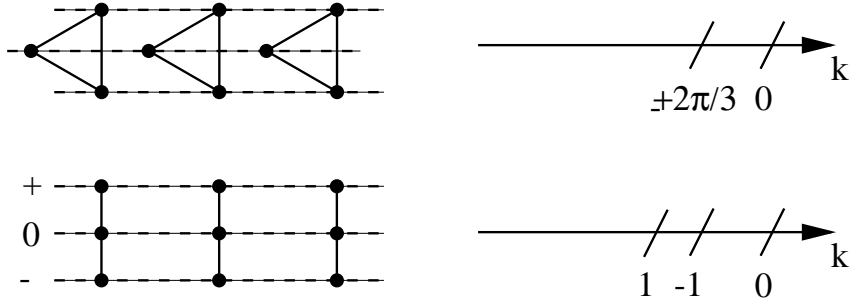

FIG. 3. Structure of three parallel chains with periodic (top) and open (bottom) boundary conditions in the transverse directions. Full and dashed lines are the perpendicular and parallel hopping matrix elements, respectively. To the right the bands crossing the Fermi energy are shown schematically. In the periodic case, the bands with $k_\perp = \pm 2\pi/3$ are degenerate. In the open case bands are labeled by their transverse quantum number, as explained in the text.

For open boundary conditions in the transverse directions, I label eigenstates on the three chains as $|-)$, $|0)$, $|+)$, as in fig.3. The eigenstates of the transverse hopping operators then are $|-1\rangle = (|+) - |-))/\sqrt{2}$, $|0\rangle = (|+) + |0) + |-))/\sqrt{3}$, and $|+1\rangle = (|+) - 2|0) + |-))/\sqrt{6}$. We now have 18 independent coupling constants. However, the renormalization group equations show that the antisymmetric $|-1\rangle$ states nearly decouple from the $|0\rangle$ and $|+1\rangle$ modes, i.e. the problem basically reduces to that of uncoupled single– and double–chain systems. For $g_1 = 0$ one then is in a $C2S2\frac{1}{2}$ state with Ising criticality, whereas for $g_1 > 0$ one has a $C2S1$ state. The dominant correlations are a form of interchain pairing involving mainly particles in the $|0\rangle$ and $|+1\rangle$ states. The decay exponent is $1/(3K_{\rho 0}) + 1/(6K_{\rho 1})$, approximately $1/2$ for weak interaction Here $K_{\rho 0}$ refers to the total charge mode and $K_{\rho 1}$ to the other massless charge mode. There are also subdominant SDW fluctuations coming from the $|-1\rangle$ transverse states, with exponent $1 + 1/(3K_{\rho 0}) + 2/(3K_{\rho 1})$, close to 2 for weak interaction. The number of massless modes for $g_1 \neq 0$ have also been determined by Arrigoni. [64]

The above is based on an analysis of the perturbative renormalization group equations and therefore valid for weak interactions. For strong intrachain interactions one can proceed similarly to the two–chain case: for $K_\rho < 1/3$ the dominant interchain interaction, generated from the initial single-particle hopping, is of the form of eq.(22). As the interaction constants renormalize to strong coupling, the semiclassical solution minimizing this term is $\phi_{\rho j} - \phi_{\rho k} \approx 0$. After the unitary transformation $\phi_j = \phi_{\sigma j}/\sqrt{2}$, $\theta_j = \sqrt{2}\theta_{\sigma j}$ $H_{jk}$ then takes the same form as the interchain coupling term of the coupled spin chain model, eq.(17), and the analysis of that Hamiltonian can be taken over. Thus, for an odd number $N$ of chains (and in particular for $N = 3$) one has one massless spin mode, and in addition a massless charge mode representing in–phase oscillations of all chains, the state thus is C1S1. The dominant fluctuations are of SDW type which decay as $r^{-1-K_\rho/N}$. On the other hand, for all even $N$ spin excitations have a gap that decreases with increasing $N$, and the state is labeled C1S0. Similar conclusions can be reached analyzing t–J ladders in the limit of large interchain exchange.

The analysis of the preceding paragraph applies as long as the interchain exchange interaction is unfrustrated, e.g. as far as $N = 3$ is concerned to open boundary conditions. For the case of periodic boundary conditions in the transverse directions the interchain exchange still scales to strong coupling, however the resulting model has quite different behavior: as far as the spin degrees of freedom are concerned, one essentially recovers a Heisenberg model on the lattice shown in the top part of fig.3, with exchange constants $J_\perp$ within each triangle and $J$ between triangles. In the limit $J_\perp \gg J$, towards which the renormalization group scales, this model can be further transformed: in the ground state, each triangle is expected to be in one of the spin–1/2 states

$$|\uparrow R\rangle = \frac{1}{\sqrt{3}}(|\downarrow\uparrow\uparrow\rangle + \omega|\uparrow\downarrow\uparrow\rangle + \omega^2|\uparrow\uparrow\downarrow\rangle) \ , \quad |\uparrow L\rangle = \frac{1}{\sqrt{3}}(|\downarrow\uparrow\uparrow\rangle + \omega^2|\uparrow\downarrow\uparrow\rangle + \omega|\uparrow\uparrow\downarrow\rangle) \ ,$$

$$|\downarrow R\rangle = \frac{1}{\sqrt{3}}(|\uparrow\downarrow\downarrow\rangle + \omega|\downarrow\uparrow\downarrow\rangle + \omega^2|\downarrow\downarrow\uparrow\rangle) \ , \quad |\downarrow L\rangle = \frac{1}{\sqrt{3}}(|\uparrow\downarrow\downarrow\rangle + \omega^2|\downarrow\uparrow\downarrow\rangle + \omega|\downarrow\downarrow\uparrow\rangle) \ , \quad (23)$$

where $\omega = \exp(2\pi i/3)$. States with total spin 3/2 on a triangle are by $J_\perp$ higher in energy and thus are neglected. In this subspace, the effective Hamiltonian becomes

$$H_{\rm eff} = \frac{J}{3}\sum_i \boldsymbol{S}_i \cdot \boldsymbol{S}_{i+1}[1 + 2(\tau_i^x \tau_{i+1}^x + \tau_i^y \tau_{i+1}^y)] \ , \qquad (24)$$

where $i$ label triangles, $\boldsymbol{S}_i$ is a spin–1/2 operator acting on the first indices of the states (23), and the $\tau_i$ are Pauli matrices acting on the second index in (23). The Hamiltonian (24) can now be analyzed using the Jordan–Wigner transformation. In the fermionic language there then are both umklapp and backward scattering process that lead to gaps in all excitation, i.e. *all the spin excitations are gapped.* For the original three–chain problem then only the total charge mode is left, the state can be labeled as C1S0. Interestingly, this is the same situation as that found in the preceding weak coupling analysis. Note also that we have here a counterexample to the general discussion of sec.II C, namely, a three–chain system with a (frustration–induced) spin gap.

Concluding the analysis of the three chain problem we may notice that, similarly to the two–chain case, for weak coupling we find dominant d–wave (or interchain pairing) fluctuations. However, for stronger intrachain interactions at least in the case of nonfrustrating boundary conditions in the transverse directions, the behavior is quite different from weak coupling, showing in particular dominant SDW fluctuations. This is quite different from the "universal" properties of the two–chain case. The principal reason seems to lie in the different mechanisms for interchain coupling: for weak interaction single–particle tunneling dominates, whereas for stronger interaction (induced) particle–hole process are more important.

## IV. CONCLUSIONS

In this paper I've discussed a number of results, mainly analytical, on the effect of different forms of interchain coupling on the Luttinger liquid behavior of strictly one–dimensional systems. As far as spin chains are concerned, the most spectacular result is the "oscillation" between even and odd numbers of chains, reminiscent of (and formally related) the Haldane phenomenon in spin–S antiferromagnetic chains. There is both experimental and numerical evidence for this behavior. For conducting chains, the situation seems to be more complicated: for very strong intrachain repulsion, similar "oscillations" as for the spin chains are predicted, however, the number of massless modes also depends on the strength of the interaction. At least for weak interaction, there are dominant $d$–type pairing fluctuations both for two and three chains. Numerical work on the two-chain system is mostly in agreement with the analytical results. These results in principle apply directly to doped spin ladders, on which experimental results however are rather scarce at the moment, and to few–channel quantum wires. In both cases are more detailed understanding of the effect of disorder is needed.

## ACKNOWLEDGMENTS

I'm grateful to T. Giamarchi, E. Orignac, R. Noack, L. Balents, and M. P. A. Fisher for stimulating discussions.

## APPENDIX A: WHEN TO BOSONIZE IN PEACE

Using eq.(7) and its generalization to cases with spin and other "internal" degrees of freedom like perpendicular momentum indices in coupled chain problems, a typical fermion interaction term becomes

$$\psi_\alpha^\dagger \psi_\beta^\dagger \psi_\gamma \psi_\delta = \eta_\alpha \eta_\beta \eta_\gamma \eta_\delta \times \text{(boson operators)} = h_{\alpha\beta\gamma\delta} \times \text{(boson operators)} \ , \tag{A1}$$

where the second equality defines $h_{\alpha\beta\gamma\delta}$. This operator, responsible for taking into account fermion anticommutation properly, is nevertheless carefully passed under the rug in the vast majority of the literature, thus leaving a purely bosonic Hamiltonian to be considered, as implied in the term "bosonization". I will here investigate under which conditions this is allowed.

The relevant situation is that all the indices $\alpha, \beta, \gamma\, \delta$ in $h_{\alpha\beta\gamma\delta}$ are different from each other. Otherwise the anticommutation rule

$$[\eta_r, \eta_s]_+ = 2\delta_{r,s} \Rightarrow \eta_r^2 = 1 \ , \tag{A2}$$

allows to simplify $h_{\alpha\beta\gamma\delta}$. I will therefore only consider the general case. First note that

$$h_{\alpha\beta\gamma\delta}^2 = 1 \ , \tag{A3}$$

$h_{\alpha\beta\gamma\delta}$ thus has eigenvalues $\pm 1$. Secondly,

$$[h_{\alpha\beta\gamma\delta}, h_{\kappa\lambda\mu\nu}]_\pm = 0 \ , \tag{A4}$$

where according to whether an even or odd number of pairs of indices taken from the two sets $(\alpha, \beta, \gamma\, \delta)$ and $(\kappa, \lambda, \mu, \nu)$ are equal the commutator (even case) or anticommutator (odd case) is to be used. Finally, permutation of indices leads to sign changes:

$$h_{\alpha\beta\gamma\delta} = -h_{\beta\alpha\gamma\delta} = -h_{\alpha\gamma\beta\delta} = -h_{\alpha\beta\delta\gamma} \tag{A5}$$

It is now clear that if all the $h$'s occuring in a given Hamiltonian commute, they can be simultaneously diagonalized, which means that it will be possible to replace each of the $h$'s by $\pm 1$, leading to a purely bosonic Hamiltonian. This clearly is the case if all the $h$'s occuring have an even number of indices in common. In the opposite case some of the $h$'s do not commute, therefore can not be simultaneously diagonalized and not be eliminated from the Hamiltonian. Bosonization then is not possible.

As a simple example consider the single–chain Luttinger model with spin, sec.III. the four allowed values of the discrete indices are $1 \equiv (+,\uparrow), 2 \equiv (+,\downarrow), 3 \equiv (-,\uparrow), 4 \equiv (-,\downarrow)$. Consequently, only $h_{1234}$ can occur, and according to the eigenvalue chosen the backward scattering interaction takes the form $\pm g_1 \cos(\sqrt{8}\phi_\sigma)$. The choice of eigenvalue of $h_{1234}$ affects however the expressions for correlation functions: for example $h_{1234} = \pm 1$ implies $\eta_1 \eta_3 = \pm \eta_2 \eta_4$, and consequently the $2k_F$ charge density operator contains either a factor $\cos(\sqrt{2}\phi_\sigma)$ (plus sign) or $\sin(\sqrt{2}\phi_\sigma)$. A similar discrete "gauge covariance" exists of course for all correlation functions.

In more complicated cases like the two–chain problem, more then one $h$–operator occurs. Even if they all commute, as is the case for the two–chain problem, additional constraints on the permissible eigenvalues of the $h$'s exist due to the existence of relations of the type

$$h_{\alpha\beta\gamma\delta} h_{\kappa\lambda\mu\nu} h_{\pi\rho\sigma\tau} = \pm 1 \ , \tag{A6}$$

and similar relations involving more than three $h$'s. However, a discrete gauge freedom of the type mentioned above often remains. For the particular case of fermions with an internal SU(N) symmetry, [9] bosonization can be performed without problem and all the $h$–operators can be given eigenvalue $+1$, a fact not noticed in the original work.

[1] D. Pines and P. Nozières, *The Theory of Quantum Liquids* (Addison–Wesley, Menlo Park, 1966).
[2] V. J. Emery, in *Highly Conducting One-Dimensional Solids*, edited by J. T. Devreese, R. P. Evrard, and V. E. van Doren (Plenum, New York, 1979), p. 247.
[3] J. Sólyom, Adv. Phys. **28**, 209 (1979).
[4] H. J. Schulz, in *Mesoscopic Qunatum Physics, Les Houches, Session LXI, 1994*, edited by E. Akkermans, G. Montambaux, J.-L. Pichard, and J. Zinn-Justin (Elsevier, Amsterdam, 1995), p. 533.
[5] J. M. Luttinger, J. Math. Phys. **4**, 1154 (1963).
[6] D. C. Mattis and E. H. Lieb, J. Math. Phys. **6**, 304 (1965).
[7] A. Luther and I. Peschel, Phys. Rev. B **9**, 2911 (1974).
[8] D. C. Mattis, J. Math. Phys. **15**, 609 (1974).
[9] T. Banks, D. Horn, and H. Neuberger, Nucl. Phys. B **108**, 119 (1976).
[10] F. D. M. Haldane, J. Phys. C **14**, 2585 (1981).
[11] H. A. Bethe, Z. Phys. **71**, 205 (1931).
[12] L. D. Faddeev and L. A. Takhtajan, Phys. Lett. A **85**, 375 (1981).
[13] P. Jordan and E. Wigner, Z. Phys. **47**, 631 (1928).
[14] I. Affleck, D. Gepner, T. Ziman, and H. J. Schulz, J. Phys. A **22**, 511 (1989).
[15] T. Giamarchi and H. J. Schulz, Phys. Rev. B **39**, 4620 (1989).
[16] R. R. P. Singh, M. E. Fisher, and R. Shankar, Phys. Rev. B **39**, 2562 (1989).
[17] F. D. M. Haldane, Phys. Rev. Lett. **45**, 1358 (1980).
[18] H. J. Schulz, Phys. Rev. B **34**, 6372 (1986).
[19] D. A. Tennant, T. G. Perring, R. A. Cowley, and S. E. Nagler, Phys. Rev. Lett. **70**, 4003 (1993).
[20] D. A. Tennant, R. A. Cowley, S. E. Nagler, and A. M. Tsvelik, Phys. Rev. B **52**, 13368 (1995).
[21] A. Luther and I. Peschel, Phys. Rev. B **12**, 3908 (1975).
[22] S. K. Satija *et al.*, Phys. Rev. B **21**, 2001 (1980).
[23] D. A. Tennant *et al.*, Phys. Rev. B **52**, 13381 (1995).
[24] A. Keren *et al.*, Phys. Rev. B **48**, 12926 (1993).
[25] M. Azuma *et al.*, Phys. Rev. Lett. **73**, 3463 (1994).
[26] T. M. Rice and E. Dagotto, Science, to appear.
[27] A. Luther and D. J. Scalapino, Phys. Rev. B **16**, 1153 (1977).
[28] T. Ziman and H. J. Schulz, Phys. Rev. Lett. **59**, 140 (1987).
[29] K. Hallberg, X. Q. G. Wang, P. Horsch, and A. Moreo, preprint, cond-mat/9603082.
[30] S. R. White, R. M. Noack, and D. J. Scalapino, Phys. Rev. Lett. **73**, 886 (1994).
[31] B. Frischmuth, B. Ammon, and M. Troyer, preprint, cond-mat/9601025.
[32] F. D. M. Haldane, Phys. Rev. Lett. **50**, 1153 (1983).
[33] S. P. Strong and A. J. Millis, Phys. Rev. Lett. **69**, 2419 (1992).
[34] I. Affleck, Phys. Rev. B **41**, 6697 (1990).
[35] A. M. Tsvelik, Phys. Rev. B **42**, 10499 (1990).
[36] K. Katsumata *et al.*, Phys. Rev. Lett. **63**, 86 (1989).
[37] N. Motoyama, H. Eisaki, and S. Uchida, Phys. Rev. Lett. **76**, 3212 (1996).

[38] D. J. Scalapino, Y. Imry, and P. Pincus, Phys. Rev. B **11**, 2042 (1975).
[39] H. J. Schulz, preprint, cond-mat/9604144.
[40] H. J. Schulz, Phys. Rev. Lett. **64**, 2831 (1990).
[41] M. Ogata, M. U. Luchini, S. Sorella, and F. Assaad, Phys. Rev. Lett. **66**, 2388 (1991).
[42] V. Meden and K. Schönhammer, Phys. Rev. B **46**, 15753 (1992).
[43] J. Voit, Phys. Rev. B **47**, 6740 (1993).
[44] A. R. Goñi *et al.*, Phys. Rev. Lett. **67**, 3298 (1991).
[45] H. J. Schulz, Phys. Rev. Lett. **71**, 1864 (1993).
[46] H. J. Schulz, Int. J. Mod. Phys. B **5**, 57 (1991).
[47] P. W. Anderson, Phys. Rev. Lett. **67**, 3844 (1991).
[48] D. G. Clarke, S. P. Strong, and P. W. Anderson, Phys. Rev. Lett. **72**, 3218 (1994).
[49] C. Bourbonnais and L. G. Caron, Int. J. Mod. Phys. B **5**, 1033 (1991).
[50] D. Boies, C. Bourbonnais, and A.-M. S. Tremblay, this conference, cond-mat/9604122.
[51] M. Fabrizio, Phys. Rev. B **48**, 15838 (1993).
[52] A. M. Finkel'stein and A. I. Larkin, Phys. Rev. B **47**, 10461 (1993).
[53] H. J. Schulz, Phys. Rev. B **53**, R2959 (1996).
[54] L. Balents and M. P. A. Fisher, preprint, cond-mat/9503045.
[55] D. V. Khveshenko and T. M. Rice, Phys. Rev. B **50**, 252 (1994).
[56] H. Tsunetsugu, M. Troyer, and T. M. Rice, Phys. Rev. B **49**, 16078 (1994).
[57] E. Dagotto, J. Riera, and D. J. Scalapino, Phys. Rev. B **45**, 5744 (1992).
[58] R. M. Noack, S. R. White, and D. J. Scalapino, Phys. Rev. Lett. **73**, 882 (1994).
[59] C. A. Hayward *et al.*, Phys. Rev. Lett. **75**, 926 (1995).
[60] C. A. Hayward and D. Poilblanc, preprint, cond-mat/9509123.
[61] K. Sano, preprint, cond-mat/9601027.
[62] R. M. Noack, S. R. White, and D. J. Scalapino, preprint, cond-mat/9601047.
[63] E. Orignac and T. Giamarchi, Phys. Rev. B **53**, to appear (1996).
[64] E. Arrigoni, preprint, cond-mat/9509145.